\documentclass[pra,aps, twocolumn,showpacs]{revtex4}
\usepackage{amsmath}
\usepackage{amssymb}
\usepackage{epsfig}
\pdfoutput=1

\begin{document}

\title{Quantitative complementarity between local and nonlocal character of quantum states in a three-qubit system}

\author{Xinhua Peng$^{1,2}$}
\email{xinhua.peng@uni-dortmund.de}
\author{Jingfu Zhang$^{1}$}
\author{Jiangfeng Du$^{1,2}$}
\author{Dieter Suter$^{1}$}
\email{Dieter.Suter@physik.uni-dortmund.de}
\affiliation{$^{1}$Fachbereich Physik, Technische Universit\"{a}t Dortmund, 44221 Dortmund, Germany}
\affiliation{$^{2}$Hefei National Laboratory for Physical Sciences at Microscale and
Department of Modern Physics, University of Science and Technology of China,
Hefei, Anhui 230026, P.R. China}
\date{\today}

\begin{abstract}

Local or nonlocal character of quantum states can be quantified and is subject to
various bounds that can be formulated as complementarity relations.
Here, we investigate the local vs. nonlocal character of pure three-qubit states 
by a four-way interferometer. 
The complete entanglement in the system can be measured as the entanglement
of a specific qubit with the subsystem consisting of the other two qubits.
The quantitative complementarity relations are verified experimentally in an
NMR quantum information processor.
\end{abstract}

\maketitle

\section{Introduction}

Classical physics groups physical entities in categories that are considered to be mutually exclusive,
such as waves and particles.
However, many experimental results are incompatible with this approach, since they can not
be explained in terms of a pure wave picture or a pure particle picture.
Quantum mechanics was developed to resolve this discrepancy and
Bohr introduced the concept of complementarity\cite{Bohr:1928aa} to emphasize the different approach.
The most familiar aspect of complementarity is perhaps the wave-particle duality.
It means, e.g., that light has characteristic properties that are usually associated with particles 
but also show behavior usually associated with waves.
If we design an experiment to measure any of these properties, it can only be achieved at the cost
of losing information about the other.

Complementarity is often illustrated by means of a two-way interferometer such as Young's double-slit experiment or a Mach-Zehnder setup. Already in 1909, the interference pattern of  \textquotedblleft single photons" was observed experimentally in a double-slit interference experiment  by Taylor\cite{Taylor:1909aa} and later by Dempster and Batho \cite{Dempster:1927aa}. 
This was of course possible only because these experiments did not provide any information about the
path taken by the photons in the double-slit interferometer. 
The same effect was also observed with many other kinds of single quantum objects including electrons \cite{MAllenstedt:1959aa,Tonomura:1989aa}, neutrons \cite{Zeilinger:1988aa}, trapped ions \cite{Eichmann:1993aa}, atoms \cite{Carnal:1991aa},
and even molecules \cite{Arndt:1999aa}.

At the qualitative level, complementarity is thus a well established concept.
More recently, it was found that complementarity can be quantified 
\cite{Wootters:1979aa,Bartell:1980aa,Greenberger:1988aa,Mandel:aa,Jaeger:1995aa,Englert:1996aa,Englert:2000aa}.
For the case of the wave-particle complementarity, it is possible to formulate it in terms of the inequality
\begin{equation}
P^{2}+V^{2}\leq 1.  \label{e.vp}
\end{equation}
In this expression,  the particle-like property is quantified by the predictability $P$, 
which specifies \textit{a priori} knowledge of the path that the system will follow 
(``which-way\textquotedblright information), 
whereas the wave-like properties are quantified by the visibility $V$ of the interference fringes. 
In the case of a pure quantum state, the inequality turns into the limiting equality. 
While this wave-particle duality was considered mostly in two-path interferometers,
it is possible to generalize it to multi-path interferometers \cite{Englert:2007aa}. 

Quantitative complementarity relations exist not only for individual quantum systems,
but even more for composite systems.
In systems consisting of two quantons, some new complementarity relations were found, 
such as the complementarity relation between single and two-particle fringe visibilities\cite{Jaeger:1993aa,Jaeger:1995aa}, 
between distinguishability and visibility\cite{Englert:1996aa}, 
and between the coherence and predictability\cite{Englert:2000aa} in a quantum eraser\cite{Scully:1997aa}. 
These properties are less directly measurable, but some can be quantified,
e.g. by two-particle interferometry. 
Many of these complementarity relations have been experimentally investigated by interferometric experiments,
using a wide range of composite two-quanton systems including photons \cite{Scully:1991aa,Schwindt:1999aa,Kim:2000aa,Abouraddy:2001aa,Pryde:2004aa}, atoms\cite{Durr:1998ab, Durr:1998aa} and nuclear spins in a bulk ensemble\cite{Peng:2003aa,Peng:2005ab,Zhu:2001aa}.

In the course of the study of complementarity in composite systems, entanglement is found to be a key entry.
As a purely quantum correlation with no classical counterpart, entanglement can be used to quantify the
\textit{non-local} aspects of the composite system. 
Some progress has been achieved in this direction, such as the complementarity relations between
distinguishability and entanglement\cite{Oppenheim:2003aa}, between spatial coherence of biphoton wave functions and
entanglement \cite{Saleh:2000aa}, between local and nonlocal information \cite{Bose:2002aa}, 
and a beautiful equality between visibility, predictability and entanglement in pure two-qubit states\cite{Jakob:2003aa}. 
Additionally, some complementarity relations in \textit{n}-qubit pure
systems were found, such as the relationship between multipartite
entanglement and mixedness for special classes of \textit{n}-qubit systems%
\cite{Jaeger:2003aa}, and between the single particle properties and the \textit{n}
bipartite entanglements in an arbitrary pure state of \textit{n} qubits\cite{Tessier:2005aa}. 

In our previous paper \cite{Peng:2005ab}, we found a complementarity
relation that exists in an \textit{n}-qubit pure state:
\begin{equation}
C_{k(ij...m)}^{2}+S_{k}^{2}=1.  \label{e.nqubit}
\end{equation}
This relation implies a tradeoff between the 
\emph{local} single-particle property ($S_{k}^{2}$) whose two constituents are $P_k^{2}$ and $V_k^{2}$, and
the \emph{nonlocal} bipartite entanglement between the particle and the remainder of the system ($C_{k(ij...m)}^{2}$), defined in terms of the marginal density operator $\rho _{k}$ \cite{Rungta:2001aa,Rungta:2003aa}
\begin{equation}
C_{k(ij...m)}=\sqrt{2\left[ 1-Tr\left( \rho _{k}^{2}\right) \right]}.
\end{equation}
Moreover, a conjecture was made: the bipartite entanglement $C_{k(ij...m)}^{2}$ 
might be equal to the sum of all possible \textit{pure} multi-particle entanglement(s) connected to this particle \cite{Peng:2005ab}. 
This conjecture was proved for pure two- and three-qubit systems \cite{Peng:2005ab}. 
Therefore, measuring the bipartite entanglement $C_{k(ij...m)}^{2}$ implies that we obtain an entire entanglement (\emph{nonlocal}) connected to this
particle. Therein the simplest case with two qubits has been verified by NMR interferometry, i.e.,  $C^{2}+S_{k}^{2}=1$, where $C$ is the concurrence of a two-qubit state $\psi$ which is related to "the entanglement of formation" \cite{Wootters:1998aa}, defined by 
$$
C(\psi)= \vert \langle \psi \vert \sigma_y \otimes \sigma_y \vert \psi^{\ast} \rangle \vert.
$$
where $\sigma_y$ is the $y$ component of the Pauli operator and  $\left\vert \psi ^{\ast}\right\rangle $ is the complex conjugate of $\left\vert \psi \right\rangle $.

The question that was left open in this earlier paper is, if it is possible to test the complementarity relation 
in a system with more than two qubits.
The present paper shows an example for such an experimental test in a pure three-qubit system.
For a pure state $|\xi\rangle$ of a three-qubit system ABC, 
we use a generalized four-way interferometer to verify the complementarity relation 
\begin{equation}
C_{A(BC)}^{2}(|\xi\rangle)+S_{A}^{2} (|\xi\rangle)=1. \label{e.3qubit}
\end{equation}

This experiment uses a specific property of pure three-qubit states. 
In the next section, we will describe this property and the experimental configuration used to measure the 
quantities of Eq (\ref{e.3qubit}). 
Sec. \ref{s.transducers} adds details about the main components (transducers) in the interference experiment.
In section IV, we combine the interferometer with state preparation and readout.
Section V gives experimental details of the implementation in an NMR quantum information processor
for different classes of pure 3-qubit states and discusses the results.

\section{Experimental setup for a three-qubit system}

\label{s.theory}

\subsection{Preferred basis}

Let us express the pure state $|\xi\rangle$ of the three-qubit system ABC in the
standard basis ${|ijk\rangle}$ ($i,j,k = 0,1$):
\begin{equation}
  \vert  \xi \rangle=\sum_{i,j,k}a_{ijk}|ijk\rangle 
\label{eq.state} .
\end{equation}
The coefficients $a_{ijk}$ are normalized to 1.
If we regard the pair BC as a single object, it makes sense to
consider the concurrence $C_{A(BC)}$ between qubit A and the 
composite object consisting of the two qubits B and C. 

An interesting and unique property of a pure state $|\xi\rangle$ of the three-qubit system helps us to design an experimental scheme 
for measuring the concurrence $C_{A(BC)}$ by an interference experiment: 
The reduced density matrix $\rho_{BC}$ has at most two nonzero eigenvalues. 
Accordingly, even though the state space of BC is four dimensional, only two of those dimensions are necessary to express the state $|\xi\rangle$ of ABC \cite{Coffman:2000aa}.
Therefore, the state  $|\xi\rangle$ can always be rewritten as

\begin{equation}
  |\xi\rangle=\sum_{i,j=0}^{1}b_{ij}| i \rangle_A|\Phi_j\rangle_{BC} 
\label{eq.state1}
\end{equation}
where $|\Phi_i\rangle$ are the eigenstates with the two nonzero eigenvalues of the reduced density matrix $\rho_{BC}= Tr_{A}(| \xi \rangle \langle \xi |)$, and the real coefficients $b_{ij}$ are normalized to 1.
Therefore, we can treat A and BC, at least for the present purpose, as a
pair of qubits in a pure state.

\begin{figure}[tbh]
\begin{center}
\includegraphics[width = 0.8\columnwidth]{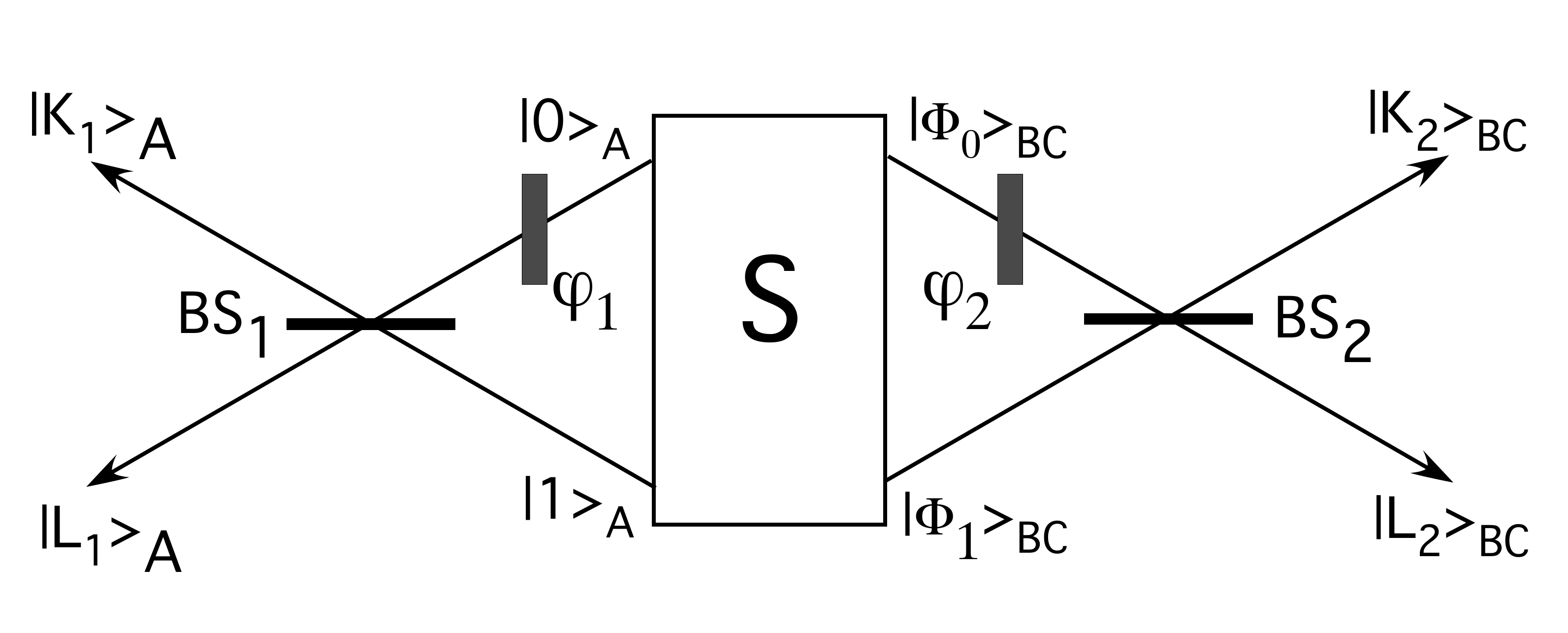}
\end{center}
\caption{Schematic four-way interferometer in a pure 3-qubit system, using beam splitters 
BS$_{1}$, BS$_{2}$ and phase shifters $\varphi _{1}$, $\varphi_{2}$.} 
\label{setup}
\end{figure}

Like in a two-qubit system, we can thus design a four-way interferometer  for a pure 3-qubit state,
which consists of four paths, which we label by the corresponding basis states 
$\{ |0\rangle_A, |1 \rangle_A, |\Phi_0\rangle_{BC}, |\Phi_1\rangle_{BC}\}$.  
Figure \ref{setup} shows the reference setup:  
The source S emits three particles A, B and C in a pure state. 
Particle A can propagate along path $|0 \rangle_A$ and/or $|1 \rangle_A$, through a variable phase shifter $\varphi_1$.
Beamsplitter BS$_1$ connects the two paths and the particles are then registered in either beam 
$|K_1 \rangle_A$ or $|L_1 \rangle_A$. 
One the other side the pair of particles B and C as a whole can propagate along the paths 
$ |\Phi_0\rangle_{BC}$ and/or $|\Phi_1\rangle_{BC}$. The probabilities of the joint (e.g. $p(|K_1 \rangle_A |K_2 \rangle_{BC})$) and single (e.g. $p(|K_1 \rangle_A)$) 
events generate interference patterns 
as a function of the phase angles $\varphi_i$, depending on the state $|\xi\rangle$ of the source S. 

From the resulting one-party interference pattern of qubit A, the single-particle fringe visibility is defined as 
\begin{equation}
\mathcal{V}_A=\frac{\left[ p\left( \left\vert x\right\rangle_A \right) \right] _{\max }-
\left[ p\left( \left\vert x\right\rangle_A \right) \right] _{\min }}{\left[
p\left( \left\vert x\right\rangle_A \right) \right] _{\max }+\left[ p\left(
\left\vert x\right\rangle_A \right) \right] _{\min }} ,
\label{def.v}
\end{equation}
where $x=K_1$ or $L_1$, and $p_{min}$ and $p_{max}$ are the minimal and maximal
probabilities (as a function of $\varphi_1 $). The other quantity related to the single-party property of particle A is the predictability 
$\mathcal{P}_A$, which quantifies the \textit{a priori} which-way knowledge. 
It is defined as  
\begin{equation}
\mathcal{P}_A =| \langle \xi | \sigma^A_z | \xi \rangle|
\label{e.predict}
\end{equation}
where $\sigma^A_z = 
\left(
\begin{array}{cc}
  1 & 0    \\
  0 & -1      
\end{array}
\right)$
is the z component of the Pauli oprator. 
$\mathcal{P}_A$ thus measures the magnitude 
of the probability difference that particle A takes path $|0 \rangle$ or the other path $|1 \rangle$.

We combine these two single-party  properties into a single entity
\[
\mathcal{S}^2_A =\mathcal{V}^2_A+ \mathcal{P}^2_A ,
\]
 which measures the single-particle character for particle A.

The two-party (\textit{nonlocal})  properties between qubit A and the pair of qubits B and C can be measured by higher order correlations.
Following references \cite{Jaeger:1995aa,Jaeger:1993aa}, we use the \textquotedblleft
corrected\textquotedblright two-party fringe visibility 
\begin{equation}
\mathcal{V}_{A(BC)} (\vert \zeta \rangle) =
\frac{\left[ \overline{p}(\vert \zeta \rangle) \right] _{\max }-\left[ \overline{p}(\vert \zeta \rangle)
\right] _{\min }}{\left[ 
\overline{p}(\vert \zeta \rangle) 
\right] _{\max }+\left[ \overline{p}(\vert \zeta \rangle) \right] _{\min }},
\label{def.v12}
\end{equation}
where the state $\vert \zeta \rangle$ is the product state
$\left\vert x\right\rangle _{A}|y\rangle _{BC}$ with 
$x=K_1$ or $L_1$ and $y=K_2$ or $L_2$. 
The \textquotedblleft corrected\textquotedblright\
joint probability $\overline{p}$ is defined as
$$
\overline{p}\left( \left\vert x\right\rangle
_{A}|y\rangle _{BC}\right) =p\left( \left\vert x\right\rangle _{A}|y\rangle
_{BC}\right) -p\left( \left\vert x\right\rangle _{A}\right) p\left( |y\rangle
_{BC}\right) +\frac{1}{4} .
$$ 
This correction eliminates single-party
contributions \cite{Jaeger:1995aa,Jaeger:1993aa}. 

The single-party and two-party properties satisfy a duality relation: 
\begin{equation}
V_{A(BC)}^{2}+\mathcal{S}_{A}^{2} = 1. 
\label{e.equal}
\end{equation}
Here $|K_1 \rangle_A$, $|L_1 \rangle_A$ is an arbitrary basis in the Hilbert space $\mathcal{H}_{A}$ of particle A.
To get the equality, we have to choose a specific basis for the BC subsystem:
$|K_2 \rangle_{BC}$, $|L_2 \rangle_{BC}$ must be linear combinations of 
the two states that correspond to the nonzero eigenvalues of the reduced density operator $\rho_{BC}$.
We will refer to this basis as the preferred basis.
With this basis, the two-party visibility $V_{A(BC)}$ becomes equal to the concurrence 
$C_{A(BC)}$, i.e., 
$$
V_{A(BC)}  \equiv C_{A(BC)} = 2 \vert b_{00}b_{11} -  b_{01}b_{10}\vert.
$$
This was proved in our previous paper \cite{Peng:2005ab}. Therefore, the concurrence $C_{A(BC)}$ of the source S can be quantitatively measured by the two-party fringe visibility $V_{A(BC)}$, so as to verify the complementarity relation (1).

\subsection{Extended basis}

For most of the calculation we assume that the measurement basis for the 
$BC$ subsystem consists of two states that are within the subspace spanned
by $\{ |\Phi_0\rangle_{BC},|\Phi_1\rangle_{BC}\}$. 
It is possible to choose a different basis, and, for an unknown input state,
it is not possible to choose a basis that falls into the 
$\{ |\Phi_0\rangle_{BC},|\Phi_1\rangle_{BC}\}$ subspace.
In the general case, the paths in the $BC$ part of the interferometer
must be written as $|m \rangle = \sum_{i=0}^{3} c_i |\Phi_i \rangle$
(with normalized coefficients $c_i$). 
The \textquotedblleft corrected\textquotedblright\
joint probability is then
\[
\overline{p}\left( \left\vert 0 \right\rangle
_{A}|m\rangle _{BC}\right) = \sum_{i=0}^{3} |c_i|^2  \overline{p}\left( \left\vert 0\right\rangle _{A}| \Phi_i \rangle
_{BC}\right)
\]
and the two-party visibility becomes
\[
\mathcal{V}_{A(BC)} \left( \left\vert 0 \right\rangle_{A}|m\rangle _{BC}\right) = \sum_{i=0}^{3} |c_i|^2 V^{(i)}_{A(BC)} 
\]
with $V^{(i)}_{A(BC)}= 
\mathcal{V}_{A(BC)}  \left( \left\vert 0\right\rangle _{A}| \Phi_i \rangle_{BC}\right )$.
Since $0 \leq |c_i|^2  \leq 1$ and $0 \leq V^{(i)}_{A(BC)} \leq 1$, 
we find
$$
\min\{V^{(i)}_{A(BC)}\} \leq  
\mathcal{V}_{A(BC)} \left( \left\vert 0 \right\rangle_{A}|m\rangle _{BC}\right)
 \leq \max\{V^{(i)}_{A(BC)}\} ,
$$
i.e.
\begin{equation}
\mathcal{V}_{A(BC)}^{2}+\mathcal{S}_{A}^{2} \le 1. 
\label{e.3qubitExp}
\end{equation}
The limiting case of the limiting equality (\ref{e.equal}) is obtained if two conditions are fulfilled:
(i) the preferred basis is chosen as the measurement basis and
(ii) the transducer acting on the BC subsystem acts only on the subspace $\{ |\Phi_0\rangle_{BC},|\Phi_1\rangle_{BC}\}$.

\section{Transducers}

\label{s.transducers}

\subsection{Preferred basis}

Both parts of our interferometer (Fig. 1) contain a transducer consisting of a
variable phase and a symmetric beam splitter.
As discussed in Ref. \cite{Peng:2005ab}, this combination provides a universal
interferometer.
Mathematically, they can be described by the unitary operation $\mathcal{\tilde{U}}$, 
written in the preferred basis $\{|i\rangle_A |\Phi_i\rangle_{BC}\}$: 
\begin{equation}
\mathcal{\tilde{U}}\left( \varphi _{1},\varphi _{2}\right) =
U_A\left( \varphi
_{1}\right) \otimes \tilde{U}_{BC}\left( \varphi _{2}\right).
\label{U.op}
\end{equation}
Each transducer $U_{A}\left( \varphi
_{1}\right)$ and  $\tilde{U}_{BC}\left( \varphi _{2}\right)$ maps
the input state into an output state by the transformation: 
\begin{equation}
U \left( \varphi_i \right)= \frac{1}{\sqrt{2}}\left(
\begin{array}{cc}
e^{-i\varphi_i /2} & e^{i\varphi_i /2} \\
-e^{-i\varphi_i /2} & e^{i\varphi_i /2}
\end{array}
\right).
\label{Uk.op}
\end{equation}
Here we use $\tilde{U}_{BC}\left( \varphi _{2}\right)$ to represent the matrix expression in 
the preferred basis $\{|\Phi_0\rangle_{BC},|\Phi_1\rangle_{BC}\}$.

\subsection{Extended basis}

However, although we can regard the pair BC as a fictitious qubit spanned by the vectors $\{|\Phi_0\rangle_{BC},|\Phi_1\rangle_{BC}\}$, the practical operation 
in the experiments on the object BC is
four dimensional. This requires us to construct a four-dimensional
unitary operation $\tilde{U}_{BC}$ in an orthonormal basis $\{|\Phi_0\rangle,|\Phi_1\rangle,|\Phi_2\rangle,|\Phi_3\rangle\}$, in whose subspace
$\{|\Phi_0\rangle,|\Phi_1\rangle\}$ the transformation has the effect of
$\tilde{U}_{BC}\left( \varphi _{2}\right)$, 
while it acts as an arbitrary single qubit operator on the subspace  
$\{|\Phi_2  \rangle,|\Phi_3\rangle\}$.  
Therefore, the transformation $\tilde{U}_{BC}$ in
the basis $\{|\Phi_0\rangle,|\Phi_1\rangle,|\Phi_2\rangle,|\Phi_3\rangle\}$ can be written in the form
\begin{widetext}
\begin{equation}
\tilde{U}_{BC}( \varphi_2)= 
\left(
\begin{array}{cccc}
\frac{1}{\sqrt{2}}e^{-i\varphi_2 /2} & \frac{1}{\sqrt{2}}e^{i\varphi_2 /2} & 0 & 0 \\
-\frac{1}{\sqrt{2}}e^{-i\varphi_2 /2} & \frac{1}{\sqrt{2}}e^{i\varphi_2/2} & 0 & 0 \\
0 & 0 & \cos \frac{\gamma}{2} e^{-i\beta} & -\sin \frac{\gamma}{2} e^{-i\delta}  \\
0   & 0 & \sin \frac{\gamma}{2} e^{i\delta} & \cos \frac{\gamma}{2} e^{i\beta} 
\end{array}\right)
\end{equation}
where $\alpha,\beta,\gamma,\delta$ are real numbers. A relative simple way is
\begin{equation}
\tilde{U}_{BC}(\varphi_2)  =  \frac{1}{\sqrt{2}}\left(
\begin{array}{cccc}
e^{-i\varphi_2 /2} & e^{i\varphi_2 /2} & 0 & 0 \\
-e^{-i\varphi_2 /2} & e^{i\varphi_2/2} & 0 & 0 \\
0 & 0 & e^{-i\varphi_2 /2} & e^{i\varphi_2 /2} \\
0 & 0 & -e^{-i\varphi_2 /2} & e^{i\varphi_2 /2}
\end{array}
\right) 
\label{U_Phi2}
\end{equation}
with $\gamma = \frac{\pi}{2}$ and $\beta =-\delta= \frac{\varphi_2}{2}$.
\end{widetext}

\section{Network}

\begin{figure}[tbh]
\begin{center}
\includegraphics[width = 0.8\columnwidth]{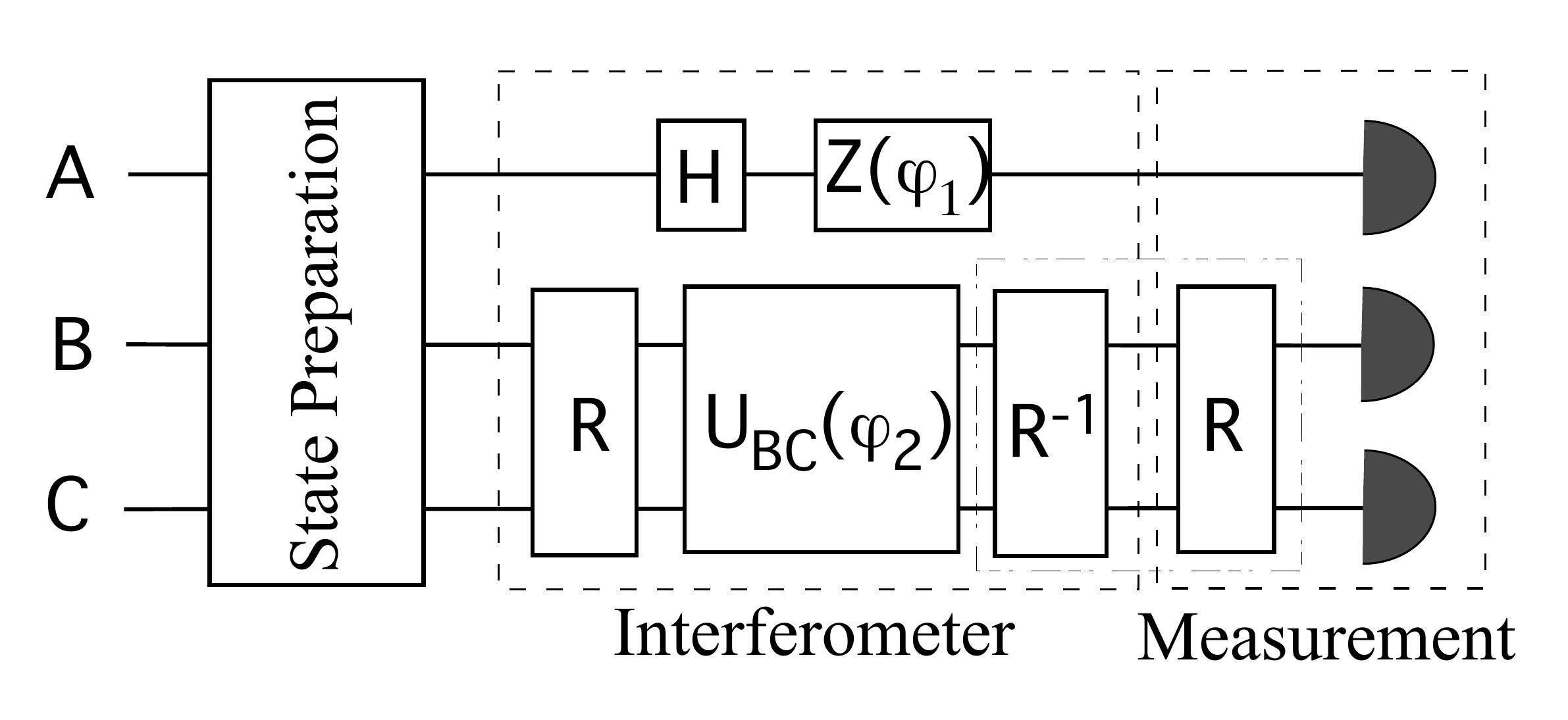}
\end{center}
\caption{Schematic network for the experimental verification of the 
complementarity relation $C_{A(BC)}^{2}+S_{A}^{2}=1$ in a pure 3-qubit system. The pseudo-Hardmard gate  $H= e^{i\frac{\pi}{4} \sigma_y}$ rotating the
qubit by the angle $\frac{\pi }{2}$ about the $-y$ axis, the phase shift gate $
Z(\varphi_1) =e^{-i\frac{\varphi_1}{2} \sigma_{z}}$ rotating the qubit 
by the angle $\varphi_1 $ about the $z$ axis, 
and the gates $\mathcal{R}$ and $U_{BC}(\varphi_2)$ are explained in the text. 
The last two operations are inverses of each other 
and can be omitted in the experiment.} 
\label{network}
\end{figure}

Fig. \ref{network} shows the network corresponding to the interferometer 
of Fig. \ref{setup} for a pure 3-qubit state. 
After the preparation of a pure 3-particle source, the transducer 
$\mathcal{\tilde{U}}_{BC}\left( \varphi _{2}\right)$ 
must be realized. 
To do this, we can use the notation: 
\[
\mathcal{\tilde{U}}_{BC}\left( \varphi _{2}\right) = \mathcal{R}^{-1} \mathcal{U}_{BC}\left( \varphi _{2}\right) \mathcal{R} 
\]
where the operation $\mathcal{R}$ transforms the chosen basis of the interferometer 
(e.g., here $\{|\Phi_0\rangle,|\Phi_1\rangle,|\Phi_2\rangle,|\Phi_3\rangle\}$)  
to the computational basis $\{|00\rangle,|01\rangle,|10\rangle,|11\rangle\}$.

The measurement observable is defined in the preferred basis  
$\{|\Phi_0\rangle,|\Phi_1\rangle\}$,
while the experimental detection scheme operates in the computational basis
$\{|00\rangle,|01\rangle,|10\rangle,|11\rangle\}$.
The dashed box labeled ``Measurement" in Fig. \ref{network} therefore starts
with a basis transformation $\mathcal{R}$, which is followed by the projective 
measurement in the computational basis.

\section{Experimental Test}

\subsection{System}

\begin{figure*}[tbh]
\begin{center}
\includegraphics[width = 1.9\columnwidth]{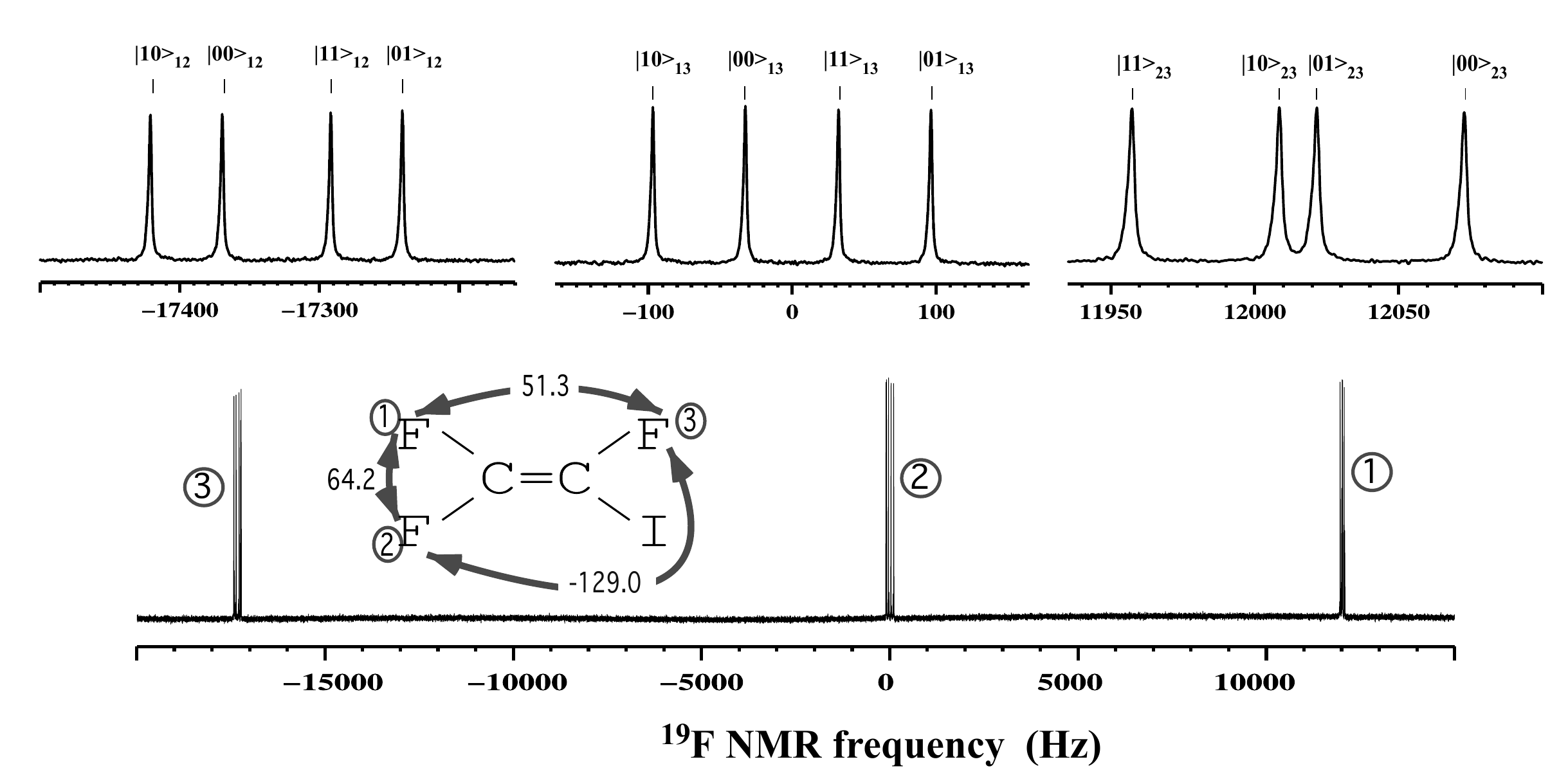}
\end{center}
\caption{$^{19}$F NMR spectrum of the Iodotrifluoroethylene, measured in a field of 11.7 T.
The chemical shifts are 12020 Hz, 0 Hz and -17330Hz.
The inset shows the structure of the molecule and the relevant coupling constants.} 
\label{sample}
\end{figure*}

As a quantum register for these experiments,  we selected the three $^{19}$F
nuclear spins of Iodotrifluoroethylene (F$_2$C=CFI), shown in the inset of Fig. \ref{sample}.
This system has relatively strong couplings between the nuclear spins,
large chemical shifts, and long decoherence times.
The Hamiltonian of this system is (in angular frequency units) 
\begin{equation}
H=\sum_{i=1}^{3}\omega _{i}I_{z}^{i}+2\pi
\sum_{i<j}^{3}J_{ij}I_{z}^{i}I_{z}^{j},  
\label{eq.H}
\end{equation}
where the $I_z^i$ are the local spin operators. Qubits 1, 2 and 3 represent particles A, B and C
of section \ref{s.theory}.

Fig. \ref{sample} shows the $^{19}$F NMR spectrum of this molecule, together with the relevant 
coupling constants.
The lower part contains the full spectrum, with the groups of lines labeled by the index
of the qubits.
The upper part shows the partial spectra of each qubit on an expanded scale.
Each qubit is coupled to the other 2 qubits, resulting in four resonance lines.
In the figure, we have labeled these lines with the corresponding logical states
of the coupled qubits.
The numerical values of the coupling constants $J_{ij}$ are given in the inset, 
together with the molecular structure.
The relaxation times are $T_1 = 5.6$ s and  $T_2 = 1.9$ s.

The experiments were performed on a Bruker Avance II 500 MHz (11.7 Tesla)
spectrometer equipped with a QXI probe with a pulsed field gradient.
The resonance frequency for the $^{19}$F spins is $\approx$ 470.69 MHz.  


\subsection{Initialization}

\begin{figure*}[tbh]
\begin{center}
\includegraphics[width = 1.9\columnwidth]{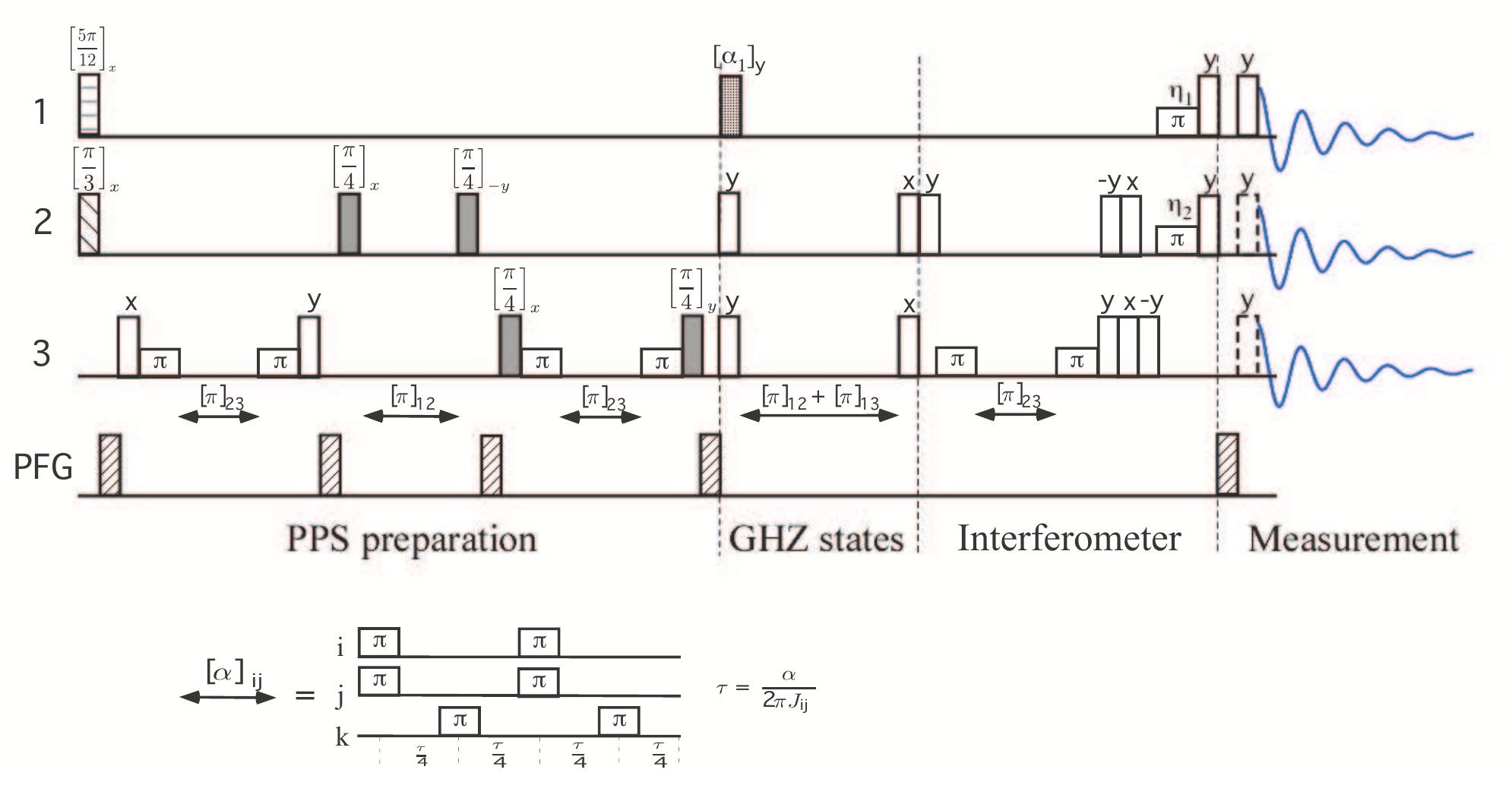}
\end{center}
\caption{(Color online) Sequence of radio-frequency and field gradient pulses (PFG) used to prepare 
the pseudopure initial state $\vert 000\rangle$ (first part), transform it into a GHZ-class state (second part), 
implement the interferometer (third part), and measure the resulting interference patterns (fourth part). 
The phases of the $\pi$ pulses in the interferometer, applied to qubits 1 and 2, are $\eta_i=\frac{\pi - \varphi_i}{2}$. $\left[ \alpha \right]_{ij}$ denotes a pure bilinear evolution of qubits $i$ and $j$  for a duration $\frac{\alpha}{2\pi J_{ij}}$, 
which is realized by the refocusing scheme shown at the bottom. } 
\label{pulse}
\end{figure*}

In the NMR experiments, the system was first prepared in a pseudopure state (PPS) 
\cite{Gershenfeld:1997aa,Cory:1997aa}
$\rho_{000} =\frac{1 - \epsilon}{8} \mathbf{1}+ \epsilon|000\rangle \langle 000|$  
with $\mathbf{1}$ representing the unity operator 
and $\epsilon \approx 10^{-5}$  the polarization,  instead of the pure state $\vert 000 \rangle$.
Starting from thermal equilibrium, we used spatial averaging \cite{Cory:1998aa} 
to prepare the PPS; the pulse sequence \cite{Peng:2002aa} is shown in the first part of Fig. \ref{pulse}.

\begin{figure}[tbh]
\begin{center}
\includegraphics[width = 0.95\columnwidth]{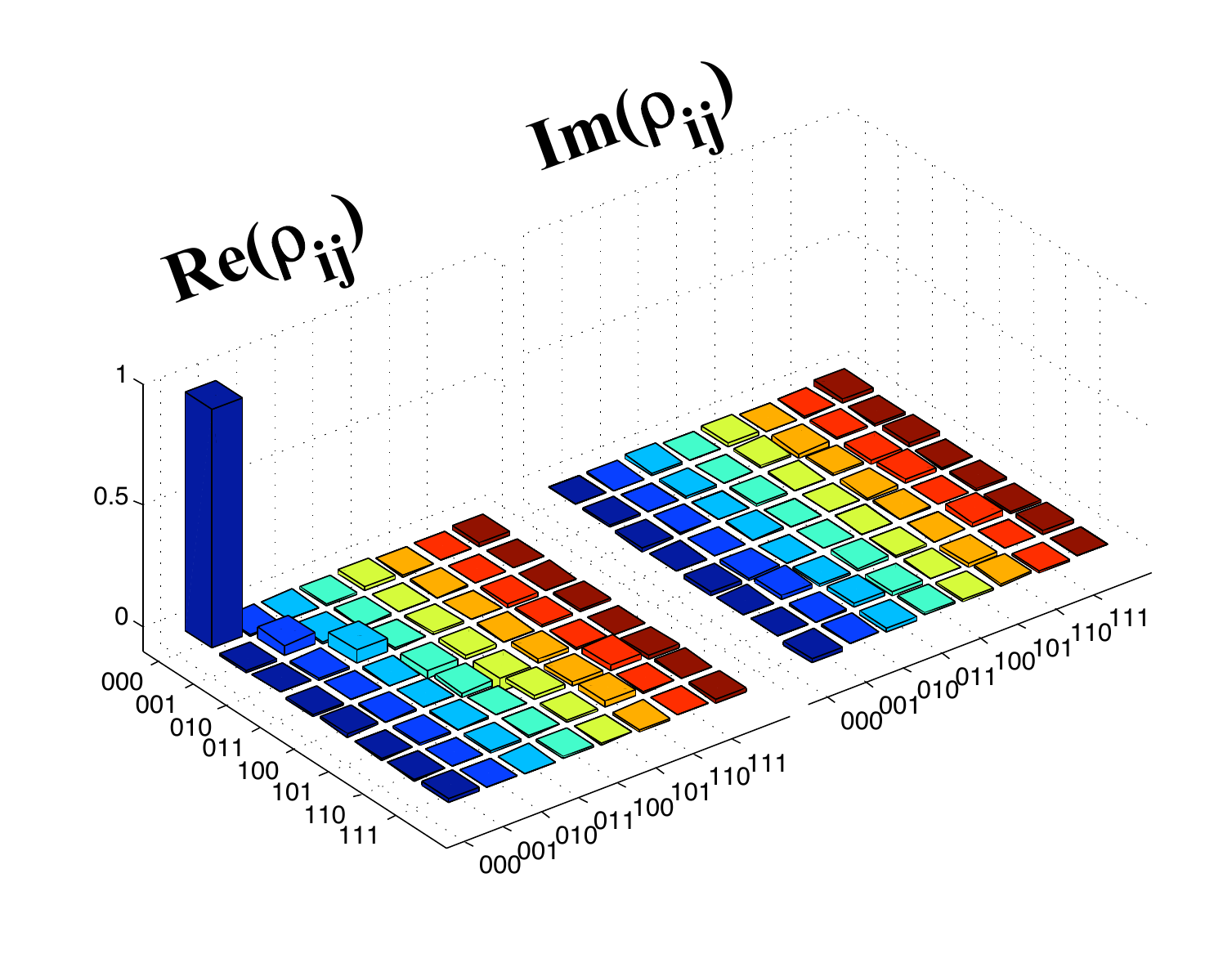}
\end{center}
\caption{(Color online) Experimentally measured relevant pure part $\rho_{exp}$ of the prepared pseudo-pure state $\rho_{000}$ reconstructed by tomography. The rows and columns are numbered with the computational basis states. } 
\label{PPS000}
\end{figure}

In the whole experiment, we used robust strongly modulating pulses 
\cite{Fortunato:2002aa,Pravia:2003aa,Mahesh:2006aa} to implement all local gates 
(e.g., $\left[ \frac{\pi}{2}\right]_{-x}^3$, and $\left[ \pi \right]_x^{1,2}$ etc.). 
In order to confirm the state preparation, we performed a complete state tomography \cite{Chuang:1998aa} to reconstruct the experimentally normalized relevant pure part $\rho_{exp}$ of the density matrix $\rho$: $\rho_{exp} \equiv \rho  - \frac{1 - \epsilon}{8} \mathbf{1}$, 
which is shown in Fig. \ref{PPS000}. 
The experimentally determined state fidelity \cite{Uhlmann:2000aa} was 
$$
F (\rho_{th},\rho_{exp}) =Tr( \sqrt{\sqrt{\rho_{th}}\rho_{exp}\sqrt{\rho_{th}}}) \approx 0.99 .
$$
With respect to scale-indenpent NMR observations and unitary evolution, a pseudo-pure state is equivalent to the corresponding pure state \cite{Vandersypen:2000ac,Das:2003aa,Laflamme.:2002aa,Ramanathan:2004aa,Negrevergne:2004aa,Laflamme:2002aa,Laflamme:1998ab,Knill:1997aa,Jones:2000ab}. 
Therefore, we focus on the relevant pure part  $\rho_{exp}$ of the pseudo-pure state in the remaining sections.

\subsection{Preparation of the 3-qubit source}

\begin{figure}[tbh]
\begin{center}
\includegraphics[width = 0.98\columnwidth]{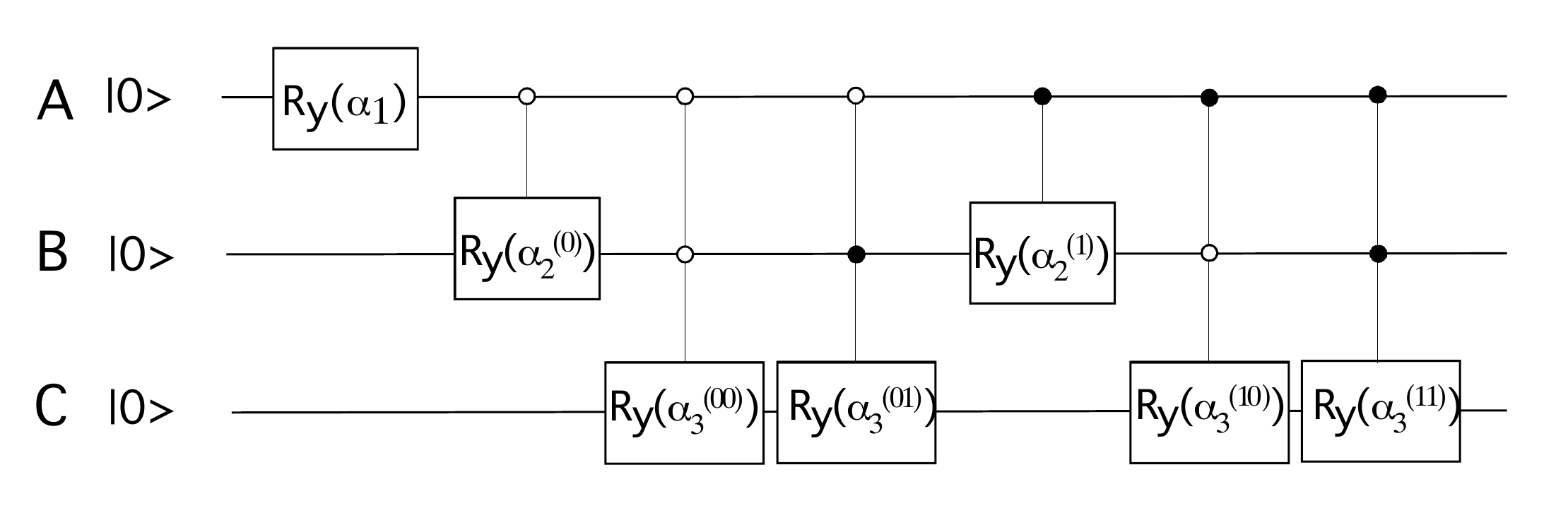}
\end{center}
\caption{Quantum circuit to prepare a pure 3-particle source state $|\xi \rangle _{ABC}$. 
Horizontal lines represent qubits; time runs from left to right. 
Conditionality on the other qubits being in the $ |1 \rangle$ and $| 0 \rangle$ state are
represented by filled and empty circles, respectively.} 
\label{state_pre}
\end{figure}

The 3-qubit states $ \vert  \xi \rangle=\sum_{i,j,k}a_{ijk} \vert ijk\rangle$ of Eq. (\ref{eq.state})
were prepared from $|000\rangle$ using the quantum circuit of Fig. \ref{state_pre}.
The resulting coefficients are
\[
a_{ijk} = \cos(\frac{\alpha_1}{2}-\frac{\pi}{2}i)\cos(\frac{\alpha^{(i)}_2}{2}-\frac{\pi}{2}j)\cos(\frac{\alpha^{(ij)}_3}{2}-\frac{\pi}{2}k).
\]

If we choose $\alpha^{(0)}_2 = \alpha^{(00)}_3 = \alpha^{(01)}_3 = \alpha^{(10)}_3 =0$ 
and $\alpha^{(1)}_2 = \alpha^{(11)}_3 = \pi$, we obtain a GHZ-class state
\begin{equation}
 \vert \xi \rangle_{GHZ} =\cos \frac{\alpha_1}{2} |000\rangle + \sin \frac{\alpha_1}{2} |111\rangle. 
\label{GHZ_state}
\end{equation}
The corresponding pulse sequence is represented in the second part of Fig. \ref{pulse}.
This type of states contains only tripartite entanglement, no bipartite entanglement.

With $\alpha^{(1)}_2 = \alpha^{(01)}_3 = \alpha^{(10)}_3  = \alpha^{(11)}_3 = 0$ and $\alpha^{(00)}_3= \pi$, we obtain a W-class state
\begin{eqnarray}
\vert \xi \rangle_{W}  & = & \cos \frac{\alpha_1}{2} \cos \frac{\alpha^{(0)}_2}{2} |001\rangle  + \nonumber \\ 
&   & \cos \frac{\alpha_1}{2} \sin \frac{\alpha^{(0)}_2}{2}  |010\rangle +\sin \frac{\alpha_1}{2}  |100\rangle.
\label{W_state}
\end{eqnarray}
These states contain only bipartite entanglement, but no tripartite entanglement.

For $\alpha^{(1)}_2 = \alpha^{(01)}_3 =  \alpha^{(10)}_3  = \alpha^{(11)}_3 =0$, we obtain the state
\begin{eqnarray}
 \vert \xi \rangle_{int} &= &\cos \frac{\alpha_1}{2} \cos \frac{\alpha^{(0)}_2}{2} \cos \frac{\alpha^{(00)}_3}{2} |000\rangle   \nonumber \\ 
&  & + \cos \frac{\alpha_1}{2} \cos \frac{\alpha^{(0)}_2}{2} \sin \frac{\alpha^{(00)}_3}{2} |001\rangle \nonumber \\
&  &+ \cos \frac{\alpha_1}{2} \sin \frac{\alpha^{(0)}_2}{2}   |010\rangle +\sin \frac{\alpha_1}{2} |100\rangle.
\label{mix_state}
\end{eqnarray}
It has tripartite as well as bipartite entanglement. 
These states are intermediate between the GHZ-class and the W-class states. 
We call them the intermediate-class states. 

\subsection{Implementation of the Interferometer}

To implement the interferometer shown in Fig. \ref{setup}, we first need to determine
the relevant eigenbasis $\{|\Phi_0 \rangle ,|\Phi_1 \rangle \}_{BC}$ of the reduced
density operator $\rho_{BC}$ of the $BC$ subsystem.
Table \ref{vecbas} lists the eigenbases for the three initial states that we consider
in this context.

\begin{table*}[htdp]
\caption{The preferred basis $\{|\Phi_0 \rangle ,|\Phi_1 \rangle \}_{BC}$ in the interferometer (see Fig. \ref{setup}) for three different classes of input states. 
For the intermediate-class states, the coefficients are $a_1 = \cos \frac{\theta_1}{2},a_2 = \sin \frac{\theta_1}{2} \sin \frac{\theta_3}{2},a_3 = \sin \frac{\theta_1}{2} \cos \frac{\theta_3}{2}$ for $|\Phi_0 \rangle$ and  $a'_1 = -\sin \frac{\theta_1}{2}, a'_2 = \cos \frac{\theta_1}{2} \sin \frac{\theta_3}{2},a'_3 = \cos \frac{\theta_1}{2} \cos \frac{\theta_3}{2}$ for $|\Phi_1 \rangle$. } 
\begin{center}
\begin{tabular}{|c|c|c|} 
\hline
State class  &  $|\Phi_0 \rangle $ & $|\Phi_1 \rangle $ \\  \hline
GHZ & $|00 \rangle$ & $|11 \rangle $ \\
W & $|00 \rangle $ & $\cos \frac{\alpha^{(0)}_2}{2} |01\rangle + \sin \frac{\alpha^{(0)}_2}{2}  |10\rangle$\\
Intermediate  & $a_1 |00\rangle +a_2  |01\rangle + a_3  |10\rangle $ & $ a'_1 |00\rangle +a'_2  |01\rangle +a'_3 |10\rangle$ \\ \hline
\end{tabular}
\end{center}
\label{vecbas}
\end{table*}

To apply the transducer in this basis, we have to find the basis transformation
between this eigenbasis and the computational basis.
Writing $\vert \Phi_0 \rangle$ as a general two-qubit state,
\begin{eqnarray}
\vert \Phi_0 \rangle & = & \cos \frac{\theta_1}{2} \cos \frac{\theta_2}{2} |00 \rangle +  \sin \frac{\theta_1}{2} \sin \frac{\theta_3}{2} |01 \rangle  \nonumber \\
&   & + \sin \frac{\theta_1}{2} \cos \frac{\theta_3}{2} |10 \rangle + \cos \frac{\theta_1}{2} \sin \frac{\theta_2}{2} |11 \rangle ,
\end{eqnarray}
we can transform it into the computational basis state $\vert 00 \rangle$
by the transformation
\begin{widetext}
\begin{eqnarray}
\mathcal{R} & = & \left(
\begin{array}{cccc}
\cos \frac{\theta_1}{2} \cos \frac{\theta_2}{2} &\sin \frac{\theta_1}{2} \sin \frac{\theta_3}{2}  & \sin \frac{\theta_1}{2}\cos \frac{\theta_3}{2}  & \cos \frac{\theta_1}{2}\sin \frac{\theta_2}{2}  \\
-\cos \frac{\theta_1}{2} \sin \frac{\theta_2}{2} &\sin \frac{\theta_1}{2} \cos \frac{\theta_3}{2}  & -\sin \frac{\theta_1}{2}\sin \frac{\theta_3}{2}  & \cos \frac{\theta_1}{2}\cos \frac{\theta_2}{2}  \\
-\sin \frac{\theta_1}{2} \cos \frac{\theta_2}{2} &\cos \frac{\theta_1}{2} \sin \frac{\theta_3}{2}  & \cos \frac{\theta_1}{2}\cos \frac{\theta_3}{2}  & -\sin \frac{\theta_1}{2}\sin \frac{\theta_2}{2}  \\
-\sin \frac{\theta_1}{2} \sin \frac{\theta_2}{2} &\cos \frac{\theta_1}{2} \cos \frac{\theta_3}{2}  & -\cos \frac{\theta_1}{2}\sin \frac{\theta_3}{2}  & \sin \frac{\theta_1}{2}\cos \frac{\theta_2}{2}  
\end{array}
\right) .
\label{R}
\end{eqnarray}
\end{widetext}
The same operator also maps the other basis states $\vert \Phi_i \rangle, i = 1, 2, 3$ 
into basis states of the computational basis. 
For the GHZ states, the pulse sequence for the implementation of this transformation is shown
in the third part of Fig. \ref{pulse}. 
Depending on the basis states $\{\vert \Phi_i \rangle \}$, a permutation of the 
computational basis states is required.

The preferred bases for the different states are shown in Table \ref{vecbas}.
For the GHZ-class states, the parameters are $\theta_i = 0, (i=1,2,3)$
and the transformation operator $\mathcal{R}$ simplifies to CNOT$_{32}$.
For the W-class states, the parameters are $\theta_1 = \theta_2 = 0, \theta_3 = - \alpha^{(0)}_2$.
For the intermediate-class states, the parameters $\theta_i$ are determined by the 
the two non-zero eigenvalues of the reduced density matrix $\rho_{BC}$,
$$
\lambda_{\pm} = \frac{1}{2} (1 \pm \sqrt{1-4A\sin ^2 \frac {\alpha_1}{2}} ,
$$
where 
$$
A = \cos ^2 \frac {\alpha_1}{2} (\cos^2 \frac{\alpha_2^{(0)}}{2} \sin^2 \frac{\alpha_3^{(00)}}{2} + \sin^2 \frac{\alpha_2^{(0)}}{2}) .
$$ 
In terms of these parameters, we find for the parameters $\theta_i$ 
in the basis transformation $\mathcal{R}$
$$
\tan^2 \frac{\theta_1}{2} = -\frac{\lambda_{-} - A}{\lambda_{+}-A}, \; \; \; \theta_2 = 0
$$ 
and 
$$
\tan  \frac{\theta_3}{2}  =\cot \frac{\alpha_2^{(0)}}{2} \sin \frac{\alpha_3^{(00)}}{2} .
$$

\subsection{Measurement}

After the interferometer, the output state of the complete 3-qubit system is 
\begin{equation}
|\psi_{out}\rangle  =  U_A (\varphi_1) \otimes \mathcal{R}^{-1} P^{\dagger}
U_{BC} (\varphi_2) P \mathcal{R} |\xi \rangle ,
\end{equation}
where $P$ is the relevant permutation matrix.
For these states, we measure the joint probabilities $p(|i \rangle_A |\Phi_j \rangle_{BC} )$
for detecting particle $A$ on port $\vert i \rangle$ and particles $B$ and $C$ on port $|\Phi_j \rangle$
of the interferometer.
This probability can be written as the projection of the output states onto the measurement basis
\begin{equation}
p(|i \rangle_A |\Phi_j \rangle_{BC} ) = | _{BC}\langle \Phi_j  | _A\langle i |   \psi_{out}\rangle |^2 .
\end{equation}
This expression can be evaluated in the computational basis $\vert ijk \rangle$
by using the transformation operator $\mathcal{R}$:
\begin{equation}
p(|i \rangle_A |\Phi_j \rangle_{BC} ) = | _{BC}\langle jk  | _A\langle i | \mathcal{R} |\psi_{out}\rangle |^2 .
\label{e.p_ijk}
\end{equation}
The single-particle probabilities of particle $A$ are
\begin{equation}
p(|i \rangle_A)  =   _A\langle i | Tr _{BC} (|\psi_{out}\rangle \langle \psi_{out} |) |i \rangle_A .
\end{equation}

\begin{figure}[tbh]
\begin{center}
\includegraphics[width = 0.98\columnwidth]{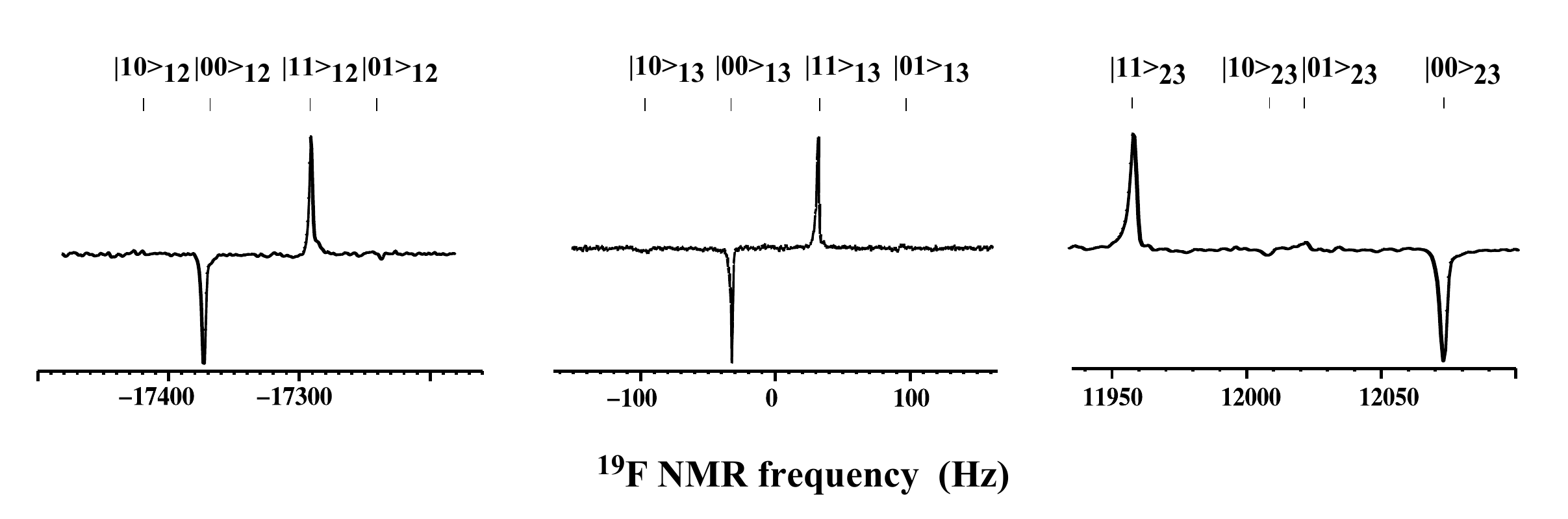}
\end{center} 
\caption{Experimental spectra of the three $^{19}$F spins (corresponding to qubits 1, 2 and 3 
from right to left) for the GHZ state when $\varphi = \varphi_1 = \varphi_2 = - \pi/2$.} 
\label{fig.spec}
\end{figure}

To determine the probabilities, we measured the populations of all eight computational basis states
by first deleting coherences with a field gradient pulse and then applying selective readout pulses to 
the individual qubits. This procedure is denoted by the last part of Fig. \ref{pulse}. The dashed read-out pulses indicate that the three pulses were applied in three separate experiments. From each of the three FIDs, we obtain a spectrum of the corresponding spin with four resonance lines after Fourier transformation. 
Fig. \ref{fig.spec} show representative spectra for the case of the GHZ state 
and a phase shift of $\varphi = \varphi_1 = \varphi_2 = - \pi /2$ in the interferometer. 

\begin{figure}[tbh]
\begin{center}
\includegraphics[width = 0.98\columnwidth]{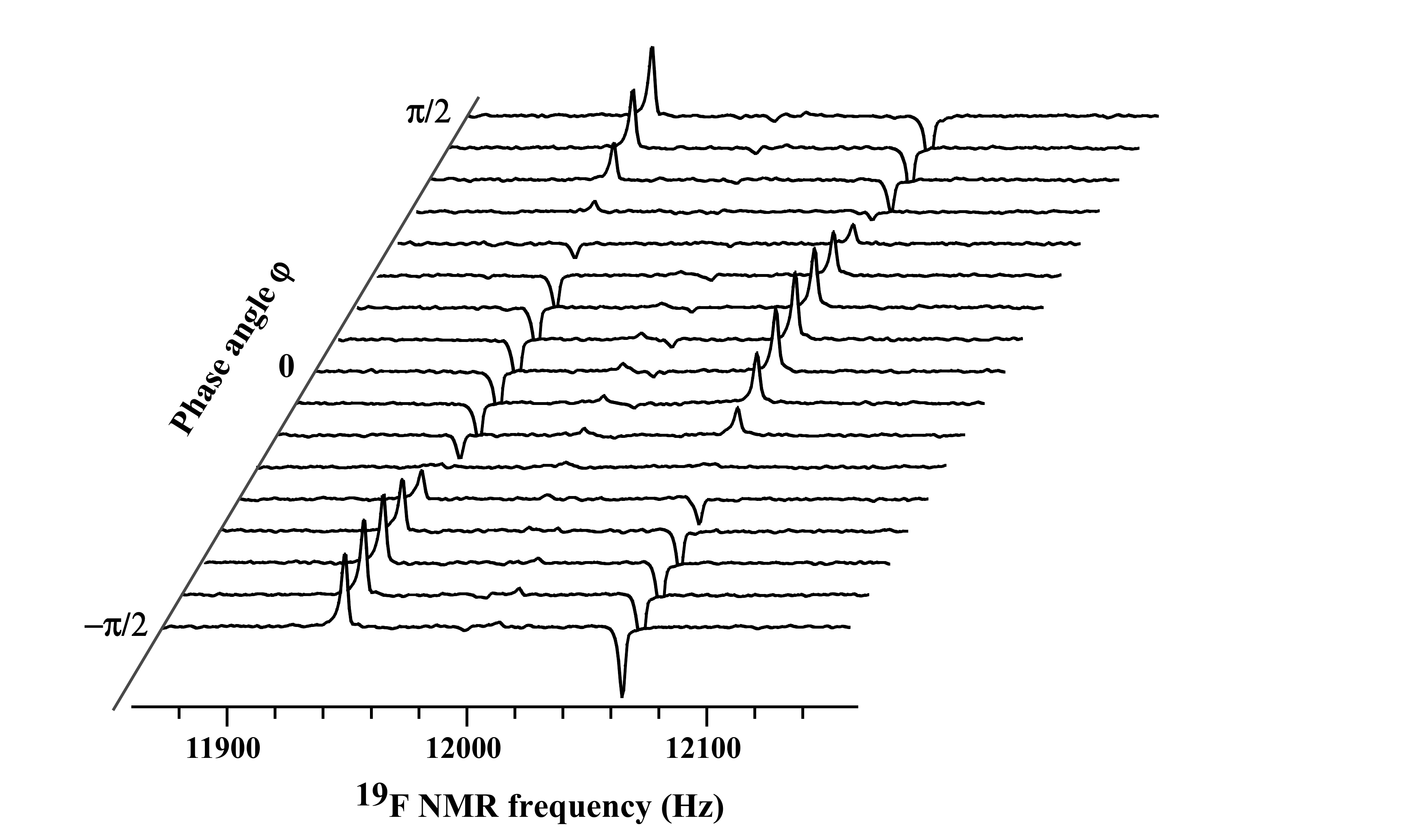}
\end{center}
\caption{Variation of the spectrum of qubit 1 with the interferometer phase 
$\varphi$ for the GHZ state.} 
\label{fig.inter}
\end{figure}

As the path length of the interferometer arms is changed,
interference between the two paths changes the populations of the different states.
This oscillation can be observed directly in the NMR spectra,
as shown in Fig. \ref{fig.inter}, where we have plotted the variation of the subspectrum of qubit 1
as a function of the phase $\varphi$ for the GHZ state.

\begin{figure}[tbh]
\begin{center}
\includegraphics[width = 0.9\columnwidth]{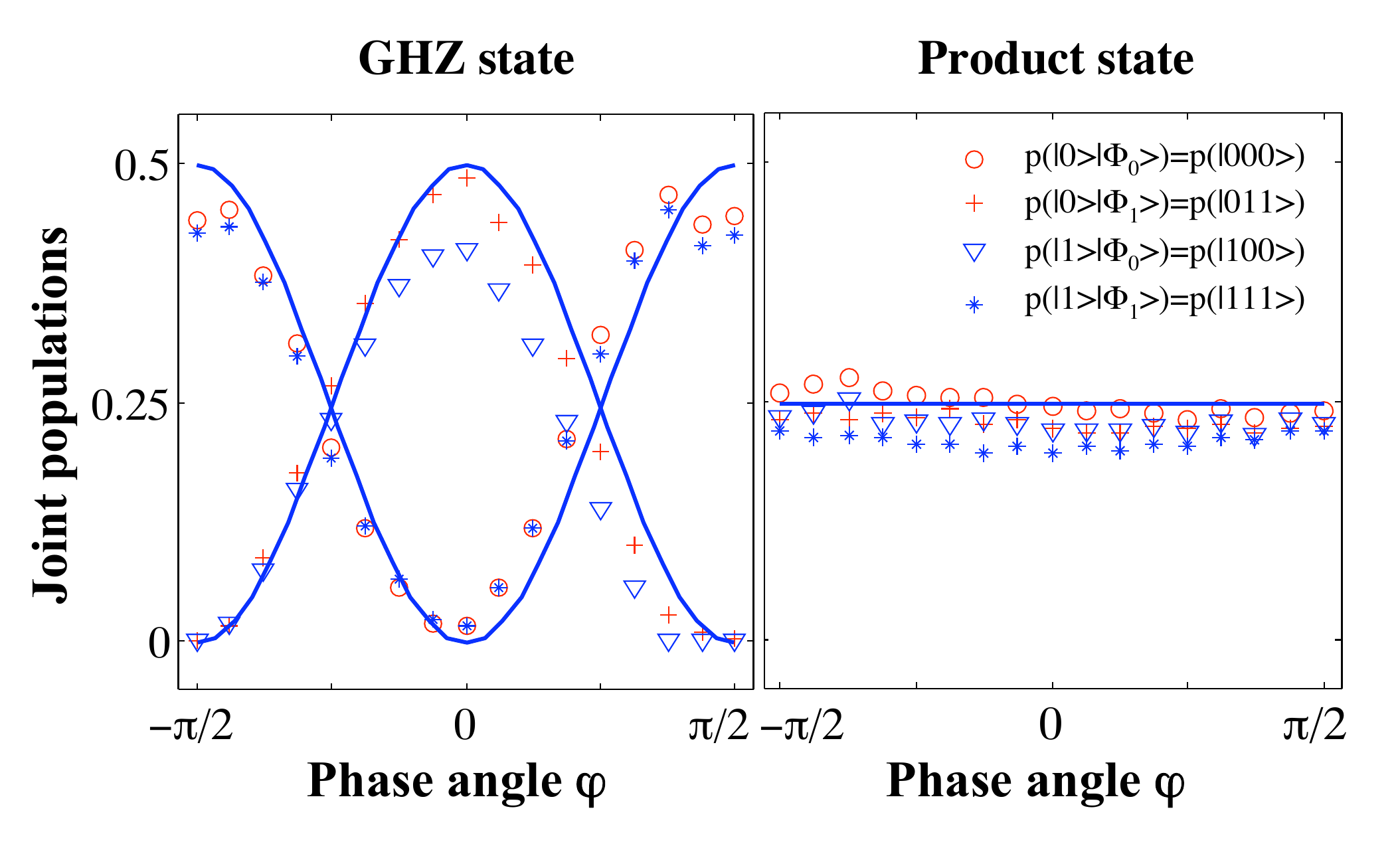}
\end{center}
\caption{(Color online) Relevant joint populations reconstructed from the experimental data for (a) the GHZ state and (b) the product state.} 
\label{fig.pop}
\end{figure}

The relevant populations were obtained from the spectra by integrating
over the resonance lines.
Fig. \ref{fig.pop} shows interferograms for some of the populations (equal to the probabilities $p(|i \rangle_A |\Phi_j \rangle_{BC}$)
for the GHZ-state (left hand side) and a product state (right hand side).
Clearly, the maximally entangled state shows high visibility fringes,
while the variation of the populations essentially vanishes for the product state. 

We extracted the visibility $V_{A(BC)}$ from the interferograms of the populations $p(|i \rangle_A |\Phi_j \rangle_{BC})$ by Eq. (\ref{def.v12}).
The single-particle probabilities $p(|i \rangle_A)$ were obtained by summing the the populations related to the state $\vert i \rangle_A$, from which we obtained the singe-particle visibility $V_A$.
The results is summarized in the next subsection.

In addition to the visibility, we measured the predictability $\mathcal{P}_{A}$.
According to Eq. (\ref{e.predict}), it is given as the expectation value of
the observable $\sigma _{z}^{(A)}$, i.e. the population difference of particle $A$.
We measured it by applying a field gradient pulse that destroyed the coherences and a
subsequent readout pulse $\left[ \frac{\pi }{2}\right] _{y}^{A}$ that converted $\sigma _{z}^{(A)}$
into $\sigma _{x}^{(A)}$, which was then recorded as the FID. 
After Fourier transformation, we integrated over the relevant resonance lines in the spectrum.
The absolute value of this integral corresponds to the predictability $\mathcal{P}_{A}$.

\subsection{Experimental results}

\begin{figure}[tbh]
\begin{center}
\includegraphics[width = 0.99\columnwidth]{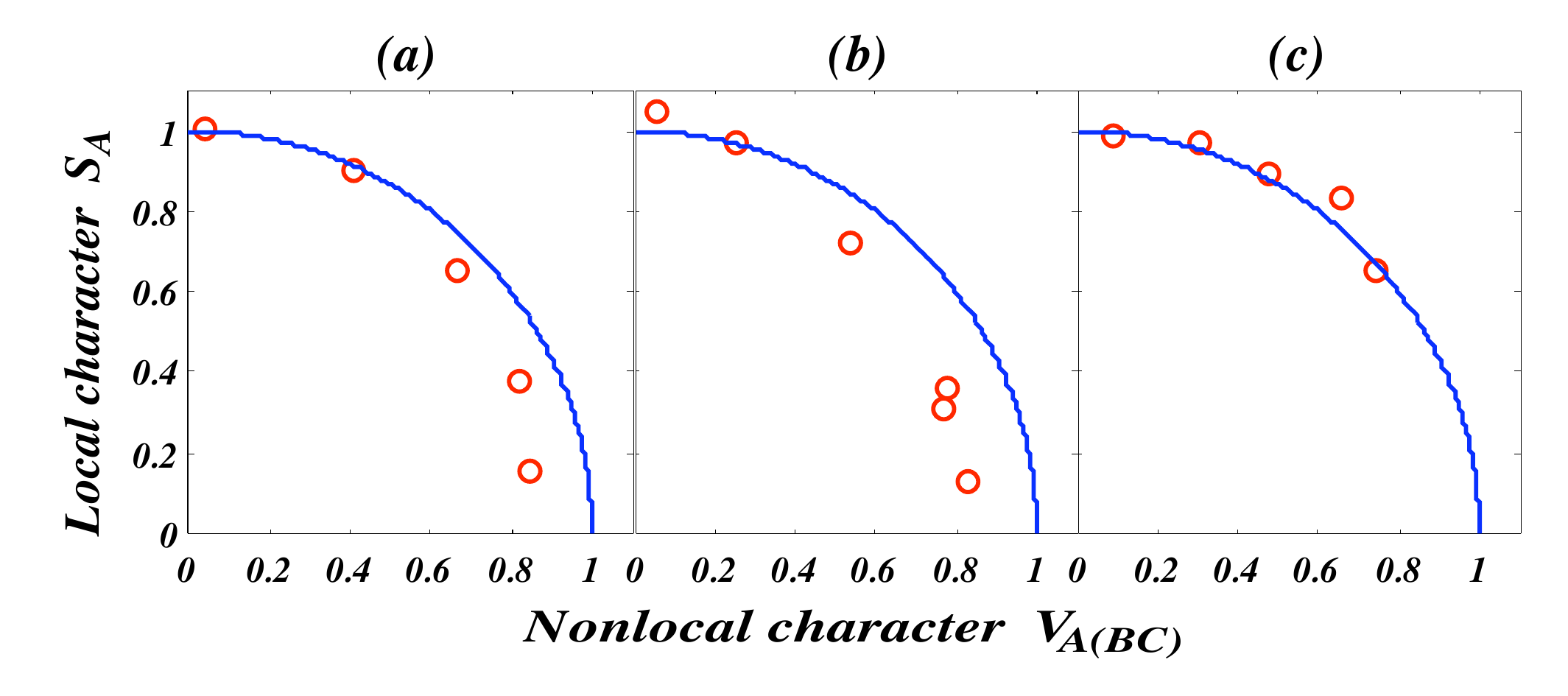}
\end{center}
\caption{(Color online) Experimental verification of the complementarity relation of $V_{A(BC)}^2 + S_A^2 = 1$ in a pure three-qubit system: (a) for GHZ-class states of Eq. (\ref{GHZ_state}), (b) for W-class states of Eq. (\ref{W_state})  and (c) for intermediate-class states of Eq. (\ref{mix_state}). Solid curves represent the theoretical complementarity relation of local character $S_A$ versus  nonlocal property $V_{A(BC)}$. Experimental results are indicated by circles. } 
\label{results}
\end{figure}

Fig. \ref{results}  summarizes the experimental results for three different classes of states
by plotting the degree of local vs. nonlocal character for each case.
According to section II, they should be related by the complementarity relation 
$V_{A(BC)}^2 + S_A^2 =1$ for the pure three-qubit states, 
where $S_A = \sqrt{V_A^2 + P_A^2}$ quantifies the local character
and $V_{A(BC)}$ the nonlocal character.
Clearly, the experimental data points (circles) agree well with the theoretical prediction (solid curves).
In the first system (GHZ), the non-local character is due exclusively to tripartite entanglement;
in the second system (W), it arises from bipartite entanglement, and in the third case,
we have a combination of both types of entanglement.

The deviation between the experimental and theoretical values is primarily due to the
inhomogeneity of the radio frequency field and the static magnetic field,
imperfect calibration of radio frequency pulses, and signal decay during the
experiments.  
The experimental errors are bigger for states with predominantly "non-local" character.
This is well compatible with the general expectation that nonlocal states are less robust
and is observed even for the initial state preparation, where the measured fidelity
is lower for the nonlocal states.
For example, the experimental fidelity of the GHZ state $\frac{1}{\sqrt{2}} (|000 \rangle +|111 \rangle) $ with $\alpha_1 = \pi/2 $ in Eq. (\ref{GHZ_state}) is about 0.97, compared to the 0.99 fidelity of the product state $|000 \rangle $ with $\alpha_1 = 0 $ in Eq. (\ref{GHZ_state}). 

\section{Conclusion}

While complementarity was first introduced as a qualitative concept,
it was recently found that in many situations, quantitative complementarity
relations can be formulated.
Here, we have quantitatively compared the local
versus nonlocal character of the quantum states of three coupled qubits (spins 1/2).

For the experimental measurements, we used a system of three coupled nuclear spins
in a liquid-state NMR spectrometer.
The degree of local vs. nonlocal character was measured by constructing a suitable four-way interferometer and utilizing a specific property of pure three-qubit states:
In any two-qubit subsystem $\rho_{BC}$, at most two eigenvectors of the density matrix
have non-zero eigenvalues. 
This allowed us to quantify the local character of the system by measuring the 
polarization of the (arbitrary) particle $A$ and the nonlocal character 
via a measurement of the entanglement between the single qubit $A$ and the subsystem $BC$.
While the interferometer only uses four channels, they were chosen in such a way that the
measurement results quantify the complete entanglement of the three-qubit system,
including bipartite as well as tripartite contributions.

Here we have restricted our theoretical treatment to cases where the three-qubit system is described in a pure quantum state. In an experiment, the prepared states inevitably involve some mixture less or more.   Naturally, the question arises as to what happens when the system is initially a mixed state. As discussed in Ref. \cite{Jakob:2003aa} and \cite{Peng:2005ab} for a bipartite system, a weaker statement for the complementarity of Eq. (\ref{e.3qubit}) is believed in the form of an inequality $C_{A(BC)}^{2}(\rho)+S_{A}^{2} (\rho) \le1$ for the most general case of a mixed three-qubit system. However, similar to the case of a bipartite system \cite{Jakob:2003aa,Peng:2005ab},  there is no corresponding inequality for the visibility $V_{A(BC)}$ in the mixed three-qubit states because the definition of $V_{A(BC)}$ by Jeager et al. \cite{Jaeger:1993aa,Jaeger:1995aa} is unfeasible and the direct relation between the concurrence $C_{A(BC)}$ and the visibility $V_{A(BC)}$ ceases to exist for mixed states.

The complementarity relation that we have verified here,
can be used to measure the degree of entanglement by measuring only
the single-particle character of a given state. 
Our measurements extend earlier tests of complementarity that were done in 
one- or two-qubit systems 
\cite{Taylor:1909aa,Mittelstaedt:1987aa,MAllenstedt:1959aa,Tonomura:1989aa,Zeilinger:1988aa, Eichmann:1993aa,Carnal:1991aa, Kim:2000aa,Scully:1991aa,Abouraddy:2001aa,Durr:1998ab, Durr:1998aa,Peng:2003aa,Peng:2005ab,Zhu:2001aa}. 
This experiment may thus be considered as a first step towards
establishing and testing quantitative complementarity relations in multi-qubit systems.

\begin{center}\textbf{ACKNOWLEDGMENTS} \par\end{center}

This work is supported by  the DFG through
Su 192/19-1, the Alexander von Humboldt Foundation and the Marie Curie program of the EU. 

\bibliography{bibliography}

\end{document}